\documentclass[conference]{IEEEtran}
\usepackage{amsmath,amsthm}
\usepackage{cite}
\usepackage{graphicx}
\usepackage{epstopdf}
\usepackage{amsfonts,amsmath,amssymb}
\usepackage{cite}
\usepackage{graphicx}
\usepackage{url}
\usepackage{bm}
\usepackage{bbm}
\usepackage{amssymb}

\begin{document}

\title{{Quantum  Entanglement Distribution in Next-Generation Wireless Communication Systems}}
\author{
\IEEEauthorblockN{Nedasadat Hosseinidehaj and Robert Malaney}
\IEEEauthorblockA{School of Electrical Engineering  \& Telecommunications,\\
The University of New South Wales,\\
Sydney, NSW 2052, Australia.\\
neda.hosseini@unsw.edu.au, r.malaney@unsw.edu.au}
}
\vspace{-1cm}
\maketitle
\begin{abstract}
In this work we analyze the distribution of quantum entanglement over  communication channels in the millimeter-wave regime. The motivation for such a study is the possibility for next-generation wireless networks (beyond 5G) to accommodate such a distribution directly - without the need to integrate additional optical communication hardware into the transceivers. Future wireless communication systems are bound to require some level of quantum communications capability. We find that direct quantum-entanglement distribution in the millimeter-wave regime is indeed possible, but that its implementation  will be very demanding from both a system-design perspective and a channel-requirement perspective.
\end{abstract}
\section{Introduction}
The exploratory design phase of Fifth Generation (5G)  wireless communications is well underway, e.g. \cite{X1,X2,X3}. These new networks will largely operate in the millimeter-wave regime (30-300GHz), where the wide bandwidths available will allow for order of magnitude improvements in data throughput. However, in order to compensate for the lack of diffraction around obstacles at high communication frequencies,  line-of-sight (LoS) channels must be utilized wherever possible.  The millimeter wavelengths used in 5G will allow for antenna technology that result in narrow beam-forming towards a LoS target receiver, with the LoS conditions being maintained (wherever possible) through location tracking of the target. Such a communication architecture represents a paradigm shift from traditional mobile wireless communication systems  in which semi-anisotropic transmissions into non-LoS communication channels are dominant.
These developments already raise interest in what  post-5G wireless networks may look like.
Here, we investigate whether millimeter-waves could also carry quantum information from a transmitter to the receiver. 
 If possible, this would greatly simplify the design of quantum-enabled post-5G wireless systems.

Traditionally, in the implementation of quantum communication protocols such as quantum teleportation  and quantum key distribution (QKD), optical frequencies have been  preferred, e.g. \cite{OP_T,OP_QKD1,OP_QKD2}. This is, in part, due to the negligible background  blackbody radiation  at optical frequencies. However, the advent of super-conducting microwave\footnote{The term \emph{microwaves} is generally used to cover the frequency range 1-30GHz.
 Henceforth, for simplicity  the term \emph{millimeter-wave} will be taken to cover the  frequency range 1-300GHz, thereby encapsulating microwaves.} quantum circuits have led to an increasing interest in the implementation of quantum communication protocols in the microwave regime, e.g. \cite{MW-QKD1, MWE1, MWE2, MWE3, MW-T1, MW-QKD2, MW-QKD3, MW-T2, MW-S, MW-R}. Such interest is further spurred by  advances in macro electro-optomechanical resonators that couple quantum information (derived from superconducting circuits) through the microwave-optical interface, e.g. \cite{MW-T1,MW-S,MW-R}. In the context of next-generation wireless communication systems, these  new developments offer the prospect of quantum information being generated at a superconducting circuit, being frequency upshifted for transfer over a wired optical network, and then being frequency-downshifted at a base station for direct transfer over millimeter-waves to a wireless receiver. In a sense, it is this paradigm we consider here. Our specific focus will be the millimeter-wave transfer of the  quantum characteristic under-pinning much of quantum communications - quantum entanglement.
 Our task appears hopeful given that previous studies have highlighted that direct QKD (no entangled states utilized) in the millimeter-wave regime is possible at low-loss rates\cite{MW-QKD2}.

To make progress we will initially consider a Gaussian  two-mode squeezed state in the millimeter-wave regime,
detailing how thermal noise affects its entanglement under photonic loss. We then consider if any advantage to entanglement transfer at these wavelengths is offered by certain non-Gaussian states.
We  also study the distribution of the  millimeter-wave entanglement between two communicating parties via an intermediate relay; using a direct transmission scheme and an entanglement swapping scheme \cite{Neda1, Neda2, Neda4}.

\section{System Model}
In the optical regime the average photon number is very low even at room temperature (300K), resulting in negligible impact of the thermal noise on the signal. In fact, in the optical frequency regime the noise variance is effectively unity, i.e. simply vacuum noise (we adopt here $\hbar=2$). However, in the millimeter-wave regime the thermal noise possesses a variance much larger than unity. In order to suppress such noise, the system needs to be operated at very low temperatures, e.g. $<100$mK.
The average photon number for a single mode is given by
$\bar n = [{{\exp \left( {hf/{k_B}T} \right) - 1}}]^{-1}\,,$
where $h$ is Planck's constant, $f$ is the frequency of the mode, $k_B$ is Boltzmann's constant, and $T$ is the temperature. A thermal quantum state, $\rho_{th}$, with average photon number $\bar n$ is a Gaussian state with zero mean and covariance matrix (CM) ${M_{th}} = \omega I$, where $\omega  = 2\bar n + 1$ is the quadrature variance and $I$ is the $2 \times 2$ identity matrix.


\subsection{Gaussian Entangled States}
We consider initially a two-mode squeezed vacuum (TMSV) state as our Gaussian entanglement resource. Production of continuous variable entanglement between millimeter-wave fields has been experimentally demonstrated \cite{MWE1, MWE2, MWE3}.
In Ref. \cite{MWE3}, the generation of TMSV states at millimeter wavelengths was experimentally demonstrated using a non-degenerate Josephson mixer. Such a mixer, driven by a pump tone at frequency $f_p$, generates a pair of entangled modes, modes~1 and 2, at different frequencies $f_1$ and $f_2$ from two input vacuum quantum states (system temperature effectively zero) such that $f_p=f_1+f_2$. Such a TMSV state with squeezing $r,\,\,r \in \left[ {0,\infty } \right)$ is a Gaussian state with zero mean and the following CM
\begin{equation}\label{M-i}
{M_v} = \left( {\begin{array}{*{20}{c}}
{vI}&{\sqrt {{v^2} - 1} Z}\\
{\sqrt {{v^2} - 1} Z}&{vI}
\end{array}} \right),
\end{equation}
where $v = cosh (2r)$ is the quadrature variance of each mode, and where $Z = diag(1, - 1)$.


Of course, the input vacuum modes can be replaced by thermal modes simply by increasing the system temperature. Doing so will lead to a thermal two-mode squeezed state as the output.
Such a Gaussian state with squeezing $r$ possesses a zero mean and a CM in the following form \cite{TTMS}
\begin{equation}\label{TMS-thermal}
\begin{array}{l}
{M_t} = \left( {\begin{array}{*{20}{c}}
{aI}&{cZ}\\
{cZ}&{bI}
\end{array}} \right),\,\,\rm where\\
\\
a = 2{{\bar n}_\alpha}{\cosh ^2}(r) + 2{{\bar n}_\beta}{\sinh ^2}(r) + \cosh (2r),\\
\\
b = 2{{\bar n}_\alpha}{\sinh ^2}(r) + 2{{\bar n}_\beta}{\cosh ^2}(r) + \cosh (2r),\\
\\
c = \left( {{{\bar n}_\alpha} + {{\bar n}_\beta} + 1} \right)\sinh (2r),
\end{array}
\end{equation}
where ${{\bar n}_\alpha}$ and ${{\bar n}_\beta}$ are the average photon numbers of the two input thermal modes~$\alpha$ and $\beta$.


\subsection{Non-Gaussian Entangled States}
We will also consider two representative non-Gaussian states, the photon-subtracted squeezed (PSS) state and the NOON state, as our non-Gaussian entanglement resource. A PSS state is created from the incoming TMSV state (with squeezing $r$) by subtracting one photon from each mode (using beam splitters with the same transmissivity $\kappa$). In terms of the Fock state ${\left| n \right\rangle}$ such a non-Gaussian state  is given by (e.g., \cite{Oxford})
\begin{equation}\label{PSSb}
\begin{array}{l}
\left| {PSS} \right\rangle  = \sum\limits_{n = 0}^\infty  {{q_{n}}} {\left| n \right\rangle _1}{\left| n \right\rangle _2},\,\\
\\
{q_{n}} = \sqrt {\frac{{{{\left( {1 - {\lambda ^2}{\kappa^2}} \right)}^3}}}{{1 + {\lambda ^2}{\kappa^2}}}} {\left( {\lambda {\kappa}} \right)^{n}}(n + 1),\\
\end{array}
\end{equation}
where $\lambda=tanh(r)$, and indices 1 and 2 indicate the two modes. The creation probability for this PSS state is given by
\begin{equation}\label{PSSb2}
\begin{array}{l}
{P} = \frac{{{\lambda ^2}\left( {1 - {\lambda ^2}} \right)\left( {1 + {\lambda ^2}{\kappa^2}} \right){{\left( {1 - {\kappa}} \right)}^2}}}{{{{\left( {1 - {\lambda ^2}{\kappa^2}} \right)}^3}}},
\end{array}
\end{equation}
and its density operator  is
\begin{equation}\label{PSSb-density}
\rho _{in}^{PSS} = \sum\limits_{m = 0}^\infty  {\sum\limits_{n = 0}^\infty  {{q_{m}}{q_{n}}{{\left| m \right\rangle }_1}{{\left\langle n \right|}_1} \otimes {{\left| m \right\rangle }_2}{{\left\langle n \right|}_2}} } .
\end{equation}
We can write the NOON state as,
$\left| {NOON} \right\rangle  = \frac{1}{{\sqrt 2 }}\left( {{{\left| n \right\rangle }_1}{{\left| 0 \right\rangle }_2} + {{\left| 0 \right\rangle }_1}{{\left| n \right\rangle }_2}} \right)$ with the following density operator
\begin{equation}\label{NOON-density}
\begin{array}{l}
\rho _{in}^{NOON} = \frac{1}{2}{\left( {{{\left| n \right\rangle }_1}{{\left\langle n \right|}_1} \otimes {{\left| 0 \right\rangle }_2}{{\left\langle 0 \right|}_2} + {{\left| n \right\rangle }_1}{{\left\langle 0 \right|}_1} \otimes {{\left| 0 \right\rangle }_2}\left\langle n \right|} \right._2}\\
\\
\left. {\,\,\,\,\,\,\,\,\,\,\,\,\,\,\,\,\,\, + {{\left| 0 \right\rangle }_1}{{\left\langle n \right|}_1} \otimes {{\left| n \right\rangle }_2}{{\left\langle 0 \right|}_2} + {{\left| 0 \right\rangle }_1}{{\left\langle 0 \right|}_1} \otimes {{\left| n \right\rangle }_2}{{\left\langle n \right|}_2}} \right).
\end{array}
\end{equation}

\subsection{Entanglement Measure}
We adopt the logarithmic negativity as our entanglement measure. The logarithmic negativity of a bipartite state with density operator $\rho $ is defined as ${E_{LN}}(\rho ) = {\log _2}\left( {1 + 2N(\rho )} \right)$, where $N(\rho )$ is the negativity. This latter quantity is defined as the absolute value of the sum of the negative eigenvalues of ${\rho ^{PT}}$, the partial transpose of $\rho $ with respect to either subsystem.
In the special case of Gaussian states, we are able to determine the logarithmic negativity solely through the use of the CM.
Given a two-mode Gaussian state with a CM
$
{M}= \left( {\begin{array}{*{20}{c}}
A&C\\
{{C^T}}&B
\end{array}} \right)
$,
where $A=A^T$, $B=B^T$, and $C$ are $2 \times 2$ real matrices, the logarithmic negativity is given by ${E_{_{LN}}}\left( {{M}} \right) = \max \left[ {0, - {{\log }_2}\left( {{\nu _ - }} \right)} \right]$, where ${\nu _ - }$ is the smallest symplectic eigenvalue of the partially transposed $M$. This eigenvalue is given by $
\nu _ - ^2 = \left( {\Delta  - \sqrt {{\Delta ^2} - 4\det M} } \right)/2
$, where $\Delta = \det A + \det B - 2\det C$ \cite{Weedbrook2012}.

\section{Simulation Results}
\begin{figure}[!t]
    \begin{center}
   {\includegraphics[width=3.3 in, height=2.2 in]{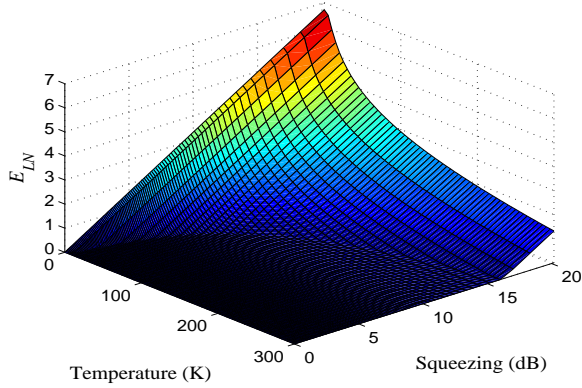}}
    \caption{The logarithmic negativity $E_{LN}$ of a thermal two-mode squeezed state as a function of temperature and squeezing for a frequency of 300GHz.}\label{thermal-3D}
    \end{center}
\end{figure}

First, we investigate the impact of the thermal fluctuations at the preparation step on the entanglement of the Gaussian two-mode squeezed state. In Fig.~\ref{thermal-3D}, we plot the logarithmic negativity, $E_{LN}$, of the thermal two-mode squeezed state, given by Eq.~\eqref{TMS-thermal}, at the frequency of 300GHz as a function of temperature and squeezing. This figure shows a setting where the two input thermal modes have the same average photon number, i.e., ${{\bar n}_\alpha} = {{\bar n}_\beta}$.

From Fig.~\ref{thermal-3D} it is evident that an increase in the temperature (an increase in the average photon number) reduces the entanglement, while an increase in the input squeezing may compensate such a negative effect. We can also see that for state-of-the-art squeezing (10dB), the temperature below which  non-zero entanglement appears is approximately 70K (interestingly, liquid nitrogen temperature). For room temperature (300K), the minimum squeezing needed to obtain the non-zero entanglement is approximately 16dB.

\subsection{Gaussian Entanglement Through a Thermal Channel}

We investigate the evolution of the  two-mode squeezed state in a thermal channel. To do this, we assume the thermal fluctuations at the preparation step can be suppressed (which means the Gaussian resource will be a TMSV state), and the thermal fluctuations only appear through  channel losses.

Let us consider Alice initially possessing a TMSV state where one mode (mode~1) is kept by Alice and the other mode (mode~2) is transmitted towards Bob over a fixed-attenuation\footnote{In the scenario we envisage, communication ranges of order 100m or less will be most relevant, and therefore neglect of attenuation fluctuations (e.g. due to multipath or beam wander) is a reasonable approximation.} channel with transmissivity $\tau$ and thermal noise variance $\omega $.
 The resultant Gaussian state is described by a zero mean and a CM in the following form
\begin{equation}\label{M-c}
{M_e} = \left( {\begin{array}{*{20}{c}}
{vI}&{\sqrt \tau  \sqrt {{v^2} - 1} Z}\\
{\sqrt \tau  \sqrt {{v^2} - 1} Z}&{\left( {\tau v + (1 - \tau )\omega } \right)I}
\end{array}} \right),
\end{equation}
where $\omega  = 2\bar n + 1$ and where $\bar n$ is the average photon number of the thermal noise.

\begin{figure}[!t]
    \begin{center}
   {\includegraphics[width=2.7 in, height=1.9 in]{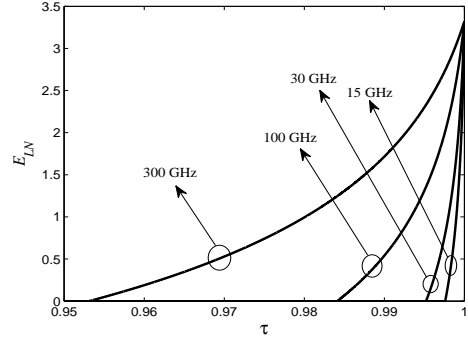}}
    \caption{The logarithmic negativity $E_{LN}$ resulting from the distribution of the TMSV state with 10dB squeezing over a single fixed-attenuation channel with transmissivity $\tau$. The figure is plotted at room temperature for different frequencies in the millimeter-wave regime.}\label{fig2}
    \end{center}
\end{figure}

In Fig.~\ref{fig2} we plot the logarithmic negativity, $E_{LN}$, of the evolved TMSV state given by Eq.~\eqref{M-c} as a function of the channel transmissivity, $\tau$. This figure shows the logarithmic negativity at different frequencies (15GHz, 30GHz, 100GHz, 300GHz) at  room temperature  when the quadrature variance of the initial TMSV states is $v=5.05$ (equivalent to 10dB squeezing). From Fig.~\ref{fig2} it is evident how, for a given $\tau$, a decrease in frequency  reduces the entanglement.
It is also evident how the entanglement-breaking transmissivity, $\tau_{eb}$, varies with frequency.
$\tau_{eb}$ for a given state is defined as the maximum transmissivity at which $E_{LN}=0$  (or, the transmissivity at which $\nu_{-} = 1$). We see from Fig.~\ref{fig2} that higher frequencies lead to lower values of $\tau_{eb}$. It is interesting to note that
for pure vacuum noise ($\omega=1$), the entanglement is always preserved for any channel transmissivity.\footnote{To be precise, the minimal background noise is actually set by the cosmic microwave background. For satellite-satellite links, assuming mK-cooled transmitter/receivers, and assuming blockage of all other background sources, $\tau_{eb}\sim 0.005$ in the millimeter-wave regime.} Whereas, for thermal noise ($\omega>1$), the entanglement disappears for $\tau \le \tau_{eb}$ where ${\tau _{eb}} = \frac{{\omega  - 1}}{{\omega  + 1}}\,$. This means that
 even though the magnitude of the entanglement at $ \tau > \tau_{eb}$ grows with increased initial squeezing, the value of  $\tau_{eb}$ itself cannot be decreased by  utilizing larger initial squeezing.

The absorption of millimeter-wave signals by the atmosphere is given in  \cite{X1}, from which we find (modulo some deep attenuations at specific frequencies) an approximately linear trend of the attenuation (in dB) as an increasing function of frequency. In the context of entanglement distribution, detrimental absorption at higher frequencies is countered to some extent by the less detrimental effects of thermalization at higher frequencies.
Assuming  atmospheric absorption is the sole  photonic loss mechanism, we find that for 30GHz at 100m, the atmospheric loss is $\sim 0.2\%$, and $E_{LN}\approx 1.4$. At 300GHz we find that at 50m, the atmospheric loss is $\sim 2.3\%$ and $E_{LN}\approx 0.8$. The entanglement-breaking distance\footnote{This is the distance at which the entanglement-breaking transmissivity is reached in the channel. Note that due to the absorption-frequency curve in the millimeter regime, lower frequencies can have larger entanglement-breaking distances than higher frequencies.} is approximately 200m and 100m for 30GHz and 300GHz, respectively. As such, we find that in principle, entanglement distribution at short distances is possible. But  given the severe loss-constraints imposed, extremely narrow millimeter-wave transmit beams must be constructed such that additional losses beyond absorption (e.g. through beam-broadening, side-lobe leakage) are constrained to the 1$\%$ (0.1$\%$) level at 300GHz (30GHz) - an interesting (\emph{aka} hard) engineering challenge.

In the classical limit (and subject to some other assumptions), power lost by beam-broadening effects is governed by the well-known Friis transmission equation
\begin{equation}\label{rob1}
P_r=P_t +G_t +G_r+20{\rm log}\left( {\frac{c_v}{{4\pi Rf}}} \right)\left[ {{\rm{dBm}}} \right] ,
\end{equation}
where $P_r$ and $P_t$ are the received (in unobstructed free space) and transmit powers, respectively;
$G_t$ and $G_r$ are the transmit and receive antenna gains, respectively; $c_v$ is the vacuum speed of light (m/s); $R$ is the transmitter-receiver distance (m), and $f$ is the signal frequency (Hz). The gains can be very much larger than unity, growing with $f^2$ for a fixed transmitter dimension. Although not \emph{directly} applicable to our problem, the Friis equation does at least highlight the magnitude of the issues we face. For example, at   the 3dB down contour, for an aperture size of $1$m and $f=300$GHz, gains of order 65dB are achieved and half-beamwidths of order 0.03$^{\rm{o}}$ obtained.\footnote{The 3dB down half-beamwidth in degrees is approximately $10/(fD)$, where  $f$ is in GHz and $D$ is the aperture size in m.} At  the 10dB down contour, for the same size and frequency, gains of order 60dB are achieved and half-beamwidths of order 0.06$^{\rm{o}}$ obtained.\footnote{See \cite{X5} for a  discussion of beam-forming in 5G.}
 Reasonable constraints on the aperture size (of say a few meters) would therefore appear to dictate that beyond transmit-receiver distances of order 100m, implementation would get very difficult. At distances of 20m or less  quantum entanglement distribution via millimeter-wave beams could be achievable for receiver aperture sizes consistent with mobile devices. Future wireless networks where quantum communications over distances of order 20m would be useful include quantum-enabled vehicular networks \cite {qcar}. 


\subsection{Non-Gaussian Entanglement Through a Thermal Channel}
We now investigate the usefulness of non-Gaussian entangled states as a resource for quantum communication in the millimeter-wave regime. Due to the unbounded number of such states  our investigation is only  indicative, not exhaustive.
Since non-Gaussian states are not completely described by the first and second moments of the quadrature operators, we cannot analyse their evolution only through the CM. Instead, we utilize the Kraus representation \cite{Kraus}. In this representation
the action of every trace-preserving completely positive channel on a quantum state with density operator ${\rho _{in}}$ can be described in an operator-sum representation of the form
${\rho _{out}} = \sum\limits_{\ell  = 0}^\infty  {{G_\ell }{\rho _{in}}\,G_\ell ^\dag }$,
where the Kraus operators ${G_\ell }$ satisfy $\sum\limits_{\ell  = 0}^\infty  {{G_\ell }\,G_\ell ^\dag }  = I$, with $I$ being the identity operator.
\begin{figure}[!t]
    \begin{center}
   {\includegraphics[width=2.7 in, height=1.9 in]{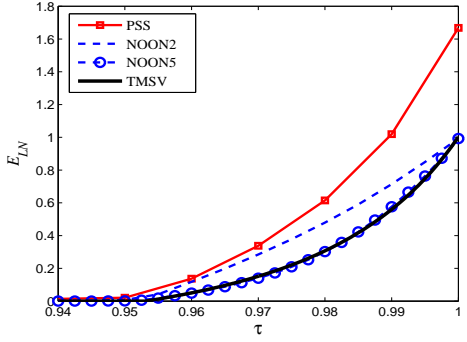}}
    \caption{The logarithmic negativity $E_{LN}$ resulting from the distribution of the TMSV state with 3dB squeezing, PSS state created from a TMSV state with 3dB squeezing, and NOON state ($n=2$ and $n=5$) over a single fixed-attenuation channel with transmissivity $\tau$ at the frequency of 300GHz and room temperature.}\label{fig3}
    \end{center}
\end{figure}
In \cite{Kraus}, the Kraus operators of a fixed-attenuation channel with vacuum noise is given.
Here, we will generalize the results of \cite{Kraus} to a fixed-attenuation channel with transmissivity $\tau$ and thermal noise ${\rho _{th}} = \sum\limits_{n = 0}^\infty  {\frac{{{{\bar n}^n}}}{{{{\left( {\bar n + 1} \right)}^{n + 1}}}}\left| n \right\rangle } \left\langle n \right|$, where $\bar n$ is the average photon number of the thermal noise. The action of such a channel on a quantum state ${\rho _{in}}$ is given by
\begin{equation}\label{thermal_Kraus}
\begin{array}{l}
{\rho _{out}} = \sum\limits_{\ell  = 0}^\infty  {\sum\limits_{n = 0}^\infty  {{G_{\ell ,n}}} } \,{\rho _{in}}\,G_{\ell ,n}^\dag ,\,\,\, \rm where\\
\\
{G_{\ell ,n}} = {\left( {\frac{{{{\bar n}^n}}}{{{{\left( {\bar n + 1} \right)}^{n + 1}}}}} \right)^{\frac{1}{2}}}\sum\limits_{m = {m_{\min }}}^\infty  {\sum\limits_{j = {j_{\min }}}^{{j_{\max }}} {\frac{{\sqrt {m!\ell !\left( {m - n + \ell } \right)!n!} }}{{\left( {m - n + j} \right)!\left( {\ell  - j} \right)!\left( {n - j} \right)!j!}}} } \\
\\
 \times {\left( { - 1} \right)^{n - j}}{\left( {\sqrt {1 - \tau } } \right)^{\ell  + n - 2j}}{\left( {\sqrt \tau  } \right)^{m - n + 2j}}\left| m \right\rangle \left\langle {m - n + \ell } \right|,
\end{array}
\end{equation}
where ${m_{\min }} = \max \left[{0,n - \ell } \right]$, ${j_{\min }} = \max \left[ {0,n - m}\right]$ and ${j_{\max }} = \min \left[ {n,\ell } \right]$.
From these operators the elementary density operator $\left| m' \right\rangle \left\langle n' \right|$ after the evolution through the thermal channel can be written
\begin{equation}\label{thermal_Kraus_elementary}
\begin{array}{l}
\left| {m'} \right\rangle \left\langle {n'} \right| \to \sum\limits_{\ell  = 0}^\infty  {\sum\limits_{n = 0}^\infty  {{G_{\ell ,n}}} } \,\left| {m'} \right\rangle \left\langle {n'} \right|\,G_{\ell ,n}^\dag  = \\
\\
\sum\limits_{n = 0}^\infty  {\sum\limits_{\ell  = 0}^{{\ell _{\max }}} {\sum\limits_{j = {j_{\min }}}^{{j_{\max }}} {\sum\limits_{j' = {{j'_{\min }}}}^{{{j'_{\max }}}} {\frac{{{{\bar n}^n}}}{{{{\left( {\bar n + 1} \right)}^{n + 1}}}}{{\left( { - 1} \right)}^{2n - j - j'}}} } } } \\
\\
 \times \frac{{\sqrt {\left( {m' + n - \ell } \right)!\ell !m'!n!} }}{{\left( {m' - \ell  + j} \right)!\left( {\ell  - j} \right)!\left( {n - j} \right)!j!}}\,\,\,\frac{{\sqrt {\left( {n' + n - \ell } \right)!\ell !n'!n!} }}{{\left( {n' - \ell  + j'} \right)!\left( {\ell  - j'} \right)!\left( {n - j'} \right)!j'!}}\\
\\
 \times {\left( {1 - \tau } \right)^{\ell  + n - j - j'}}{\left( {\sqrt \tau  } \right)^{m' + n' - 2\ell  + 2j + 2j'}}\\
\\
 \times \left| {m' + n - \ell } \right\rangle \left\langle {n' + n - \ell } \right|,
\end{array}
\end{equation}
where ${\ell _{\max }} = \min \left[ {m' + n,n' + n} \right] $, ${j_{\min }} = \max \left[ {0,\ell  - m'} \right]$, ${{j'_{\min }}} = \max \left[ {0,\ell  - n'} \right]$, and ${j_{\max }} = {{j'_{\max }}} = \min \left[ {n,\ell } \right]$.
Here, we assume for each of our non-Gaussian density operators, $\rho _{in}^{PSS}$ in Eq.~\eqref{PSSb-density} and $\rho _{in}^{NOON}$ in Eq.~\eqref{NOON-density}, mode~1 remains unaffected and mode~2 evolves through a fixed attenuation channel with transmissivity $\tau$ and thermal noise $\rho_{th}$ according to Eq.~\eqref{thermal_Kraus_elementary}. Since the logarithmic negativity, $E_{LN}$, is determined from the partial transpose of the output density operator, $\rho _{out}^{PT}$, which possesses an infinite number of elements, we are required to deploy a numerical method to approximate $\rho _{out}^{PT}$ by limiting its size, i.e. creating a truncated $\rho _{out}^{PT}$. In the case of the evolved PSS state, we will create a truncated $\rho _{out}^{PT}$ by setting two cutoffs, one cutoff on the total photon number of the two modes (see \cite{Neda3, Neda5}), and the other cutoff on the number of elements in the thermal noise $\rho_{th}$ (large relative to the average photon number of $\rho_{th}$). However, in the case of the evolved NOON state setting the latter cutoff is enough.  The logarithmic negativity can then be determined directly from the negative eigenvalues of the truncated $\rho _{out}^{PT}$.

In Fig.~\ref{fig3} we plot the logarithmic negativity $E_{LN}$ of our evolved non-Gaussian states compared to the TMSV state as a function of the channel transmissivity $\tau$. This figure shows $E_{LN}$ at a frequency of 300GHz for room temperature. Since the pure NOON states, regardless of the value of $n$, contain 1 ebit of entanglement, (i.e., $E_{LN}=1$), we consider a TMSV state with $v=1.25$ (3dB squeezing) which contains 1 ebit of entanglement. We assume the PSS state is also created from a TMSV state having an initial entanglement of 1 ebit. For the photon subtraction operation we select an optimal value of $\kappa=1$ which maximizes the initial $E_{LN}$. This means the PSS state contains an initial entanglement of more than 1 ebit. From Fig.~\ref{fig3}, we find that if the sending rates (into the channel) for all the states are equal \cite{Neda3}, some non-Gaussian states can provide an advantage relative to the Gaussian states in terms of the final entanglement. For instance, the PSS state and the NOON state with $n=2$ lead to more logarithmic negativity than the TMSV state. The `take-away' message here is that non-Gaussianity can indeed enhance entanglement distribution in the millimeter-regime, but that any enhancement is a complicated function of how  the non-Gaussian state is produced (i.e its initial entanglement), and its fragility under photonic loss. The calculations shown here  indicate that the enhancements potentially derived from non-Gaussian states may not warrant the additional complexity required for their production.

\subsection{Quantum Communication via a Relay}

Beyond the single point-to-point fixed-attenuation channel, we now turn our attention to relay-based communications. 
Specifically, we consider an entanglement swapping scheme at a relay as a means to generate  entanglement between Alice and Bob (who now have no direct LoS between them). We are interested in quantifying the \emph{cost} (reduced entanglement) incurred by such a scheme in comparison to a direct transmission (a transmission in which the relay simply forwards the signal) between Alice and Bob. Such a cost is important to determine in the context of the quantum repeater scenario - in which entanglement distillation takes place at the relay prior to any entanglement swapping operations. If this cost is too high it could be that distillation is best obtained at the end stations directly and not at the relay. For clarity of exposition, we will consider only  Gaussian states sent via the relay.


\emph{Direct Transmission Scheme:} Let us consider Alice initially possessing a TMSV state given by the CM in Eq.~\eqref{M-i}. We assume one mode is kept by Alice while the other mode is transmitted over a fixed-attenuation channel with transmissivity $\tau_a$ and thermal noise variance $\omega _a$ to an intermediate relay, then perfectly reflected in the relay and sent towards Bob through a fixed-attenuation channel with transmissivity $\tau_b$ and thermal noise variance $\omega _b$. As a result of such a transfer the Gaussian state is given by zero mean and the following CM
\begin{equation}\label{M-d}
\begin{array}{l}
{M_d} = \left( {\begin{array}{*{20}{c}}
{vI}&{\sqrt {{\tau _a}{\tau _b}} \sqrt {{v^2} - 1} Z}\\
{\sqrt {{\tau _a}{\tau _b}} \sqrt {{v^2} - 1} Z}&{v'I}
\end{array}} \right),\rm where\\
\\
v' = {\tau _b}\left( {{\tau _a}v + (1 - {\tau _a}){\omega _a}} \right) + (1 - {\tau _b}){\omega _b}.
\end{array}
\end{equation}

\emph{Entanglement Swapping Scheme:} Let us consider both Alice and Bob initially possessing a TMSV state, given by the CM in Eq.~\eqref{M-i}, with the same squeezing. One mode of each entangled state is kept by the communicating party and the second mode of each state is transmitted to the intermediate relay. We again assume Alice's (Bob's) link is a fixed-attenuation channel with transmissivity $\tau_a$ ($\tau_b$) and thermal noise variance $\omega _a$ ($\omega _b$). The received modes are then swapped via a Bell measurement at the relay (the modes are mixed through a balanced beam splitter with the two output modes  conjugately homodyned). The Bell measurement outcome is then communicated to Alice and Bob so that they can displace their modes according to the measurement outcome. Under the assumption of optimal displacement the conditional state (averaged over all possible Bell measurements) between Alice and Bob is given by the following CM \cite{Neda1, Obermaier}
\begin{eqnarray}\label{M-s}
\begin{array}{l}
{M_{s}} = \left( {\begin{array}{*{20}{c}}
{vI}&0\\
0&{vI}
\end{array}} \right) - \,\\
\\
\,\,\,\,\,\,\,\,\,\,\,\,\,\,\,\,\,\left( {{v^2} - 1} \right)\left( {\begin{array}{*{20}{c}}
{\frac{{{\tau _A}}}{\theta }I}&{ - \frac{{\sqrt {{\tau _A}{\tau _B}} }}{\theta }Z}\\
{ - \frac{{\sqrt {{\tau _A}{\tau _B}} }}{\theta }Z}&{\frac{{{\tau _B}}}{\theta }I}
\end{array}} \right),
\end{array}
\end{eqnarray}
where $\theta  = ({\tau _a} + {\tau _b})v + (1 - {\tau _a}){\omega _a} + (1 - {\tau _b}){\omega _b}$.

\begin{figure}[!t]
    \begin{center}
   {\includegraphics[width=2.7 in, height=1.9 in]{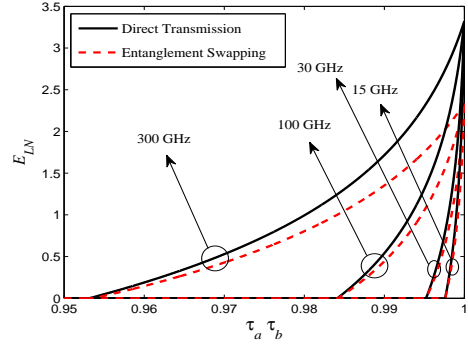}}
    \caption{The logarithmic negativity $E_{LN}$ resulting from the direct transmission scheme (solid line) and the entanglement swapping scheme (dashed line) as a function of the combined channel transmissivity $\tau_a\tau_b$. The figure is plotted at room temperature for different frequencies in the millimeter-wave regime with the initial TMSV state possessing 10dB squeezing.}\label{fig4}
    \end{center}
\end{figure}

For focus, we  consider a symmetric setting for each scheme where  Alice's link is identical to  Bob's link, i.e., ${\tau _a} = {\tau _b} = \tau$ and ${\omega _a} = {\omega _b} = \omega $.
In Fig.~\ref{fig4} we plot the logarithmic negativity resulting from our two communication schemes, as a function of the \emph{combined} channel transmissivity $\tau_a\tau_b$.  This figure shows $E_{LN}$ at different frequencies (15GHz, 30GHz, 100GHz, 300GHz) at  room temperature for the case where the quadrature variance of the initial TMSV state is $v=5.05$ (again, equivalent to 10dB squeezing). The same settings are adopted for the two schemes. Note, the curves of $E_{LN}$ for the direct transmission scheme with respect to $\tau_a\tau_b$ in Fig.~\ref{fig4} are exactly the same as the corresponding curves of $E_{LN}$ with respect to $\tau$ in Fig.~\ref{fig2}.

According to Fig.~\ref{fig4} the direct transmission scheme is always able to generate more entanglement than the entanglement swapping scheme. However, the entanglement-breaking transmissivity at each frequency is almost the same for the two schemes. Considering the symmetric setting, the entanglement-breaking transmissivity (in each channel) of the direct transmission scheme is given by $\tau _{eb}^d = \frac{{\sqrt {{\omega ^2} - 1} }}{{\omega  + 1}}$, while  for the entanglement swapping scheme it is given by $\tau _{eb}^s = \frac{\omega }{{\omega  + 1}}$. From this we see that as  $\omega$ increases, the entanglement-breaking transmissivity for the two schemes becomes identical. Note again, that the value of $\tau _{eb}$ for each scheme is only dependent on the thermal noise of the channel.

From these results it is evident that entanglement swapping at the relay in the non-LoS condition between Alice and Bob can occur at only a slight relative cost, and that direct transmission (reflection at the relay) is more useful in terms of the entanglement distribution between Alice and Bob. The necessity for entanglement distillation at the relay (as opposed to at Alice and/or Bob) would be a function of the distillation protocol (some protocols operate better at higher input entanglement levels). Of course, in a lossless channel, direct transmission is always a preferred transmission scheme.


\subsection{QKD and other Applications}
Although difficult to achieve in practice, we have seen that  non-zero entanglement distribution in the  millimeter-wave regime may be possible over short distances. This then begs the question as to what you can do with such entanglement. A first use of course is QKD. It is well known that all QKD implementations have a prepare and measure (PM) scheme  - and an entirely equivalent entanglement scheme (e.g. \cite{Weedbrook2012}). Based on this correspondence, we have carried out additional calculations which translate our entanglement distribution calculations for Gaussian states into resultant QKD rates (we refer the reader to \cite{Neda4} for more detail on such translations). The resulting QKD key-rates for direct transmissions  are entirely consistent with those found previously for a PM QKD scheme based on thermal states \cite{MW-QKD2}.  Our key rates for the entanglement swapping scheme (based on assuming no correlation between ancillary states used by the eavesdropper) are somewhat lower, as expected.  Our main conclusion on key rates is therefore is similar to that found for entanglement distribution - QKD is possible in the millimeter-wave regime but difficult and viable over only short distances.

Other quantum applications  are possible with the entanglement levels we have found (teleportation, dense coding \emph{etc.}). However, given the low values of the entanglement achievable, such applications will likely need substantial entanglement distillation prior to the application running, in which case the use of non-Gaussian states such as those investigated here would be useful. We refer the reader to \cite{Neda1} for an example of a distillation scheme in the non-Gaussian space.


\section{Conclusions}
We have shown how direct quantum entanglement distribution in next-generation wireless communications is possible  - but very demanding on the design requirements of the system and on the channel requirements of the communication links. In effect, almost pure (negligible non-LoS components) and almost lossless links will be necessary for entanglement transfer to be viable in the millimeter-wave regime. Nonetheless, entanglement distribution via millimeter communication is potentially viable in practice at communication distances of less than 100m. For distances much beyond 100m any entanglement distribution would be extremely difficult to implement in practice.

It remains an open question as to whether in practice the implementation complexity involved in the delivery of low-loss  millimeter-wave quantum links renders  a full integration of free-space optical quantum communications into wireless architectures a better option. The latter scheme will have less demands on the channel and beam characteristics,
but will necessitate that the transceivers have on board the hardware to operate in two quite different  wavelength regimes. Future work could explore this question further.


%

\end{document}